# High-Throughput In-Memory Computing for Binary Deep Neural Networks with Monolithically Integrated RRAM and 90nm CMOS

Shihui Yin[1], Xiaoyu Sun[2], Shimeng Yu[#2], and Jae-sun Seo*[1]

[1]School of Electrical, Computer and Energy Engineering, Arizona State University, Tempe, AZ, USA

[2]School of Electrical and Computer Engineering, Georgia Institute of Technology, Atlanta, GA, USA

* Correspondence to jaesun.seo@asu.edu, [#]shimeng.yu@ece.gatech.edu

Deep learning hardware designs have been bottlenecked by conventional memories such as SRAM due to density, leakage and parallel computing challenges. Resistive devices can address the density and volatility issues, but have been limited by peripheral circuit integration. In this work, we demonstrate a scalable RRAM based in-memory computing design, termed XNOR-RRAM, which is fabricated in a 90nm CMOS technology with monolithic integration of RRAM devices between metal 1 and 2. We integrated a 128×64 RRAM array with CMOS peripheral circuits including row/column decoders and flash analog-to-digital converters (ADCs), which collectively become a core component for scalable RRAM-based in-memory computing towards large deep neural networks (DNNs). To maximize the parallelism of in-memory computing, we assert all 128 wordlines of the RRAM array simultaneously, perform analog computing along the bitlines, and digitize the bitline voltages using ADCs. The resistance distribution of low resistance states is tightened by write-verify scheme, and the ADC offset is calibrated. Prototype chip measurements show that the proposed design achieves high binary DNN accuracy of 98.5% for MNIST and 83.5% for CIFAR-10 datasets, respectively, with energy efficiency of 24 TOPS/W and 158 GOPS throughput. This represents 5.6X, 3.2X, 14.1X improvements in throughput, energy-delay product (EDP), and energy-delay-squared product (ED$^2$P), respectively, compared to the state-of-the-art literature. The proposed XNOR-RRAM can enable intelligent functionalities for area-/energy-constrained edge computing devices.

# Introduction

Deep neural networks (DNNs) have been very successful in large-scale recognition and classification tasks, some even surpassing human-level accuracy[1–5]. To achieve incremental accuracy improvement, state-of-the-art deep learning algorithms tend to present very deep and large network models[1–3], which poses significant challenges for hardware implementations in terms of computation, memory, and communication. This is especially true for embedded hardware applications, such as autonomous driving[6], machine translation[7], and smart wearable devices[8], where severe constraints exist in performance, power, and area. To address this on the algorithm side, recent works aggressively lowered the precision to the extreme where both the weights and neuron activations are binarized to +1 or -1[9,10] for inference, such that the multiplication between weights and activations becomes XNOR operation and accumulation becomes bitcounting of bitwise XNOR values. Such binarized neural network (BNN) algorithms largely reduce the computational complexity and weight memory requirement.

On the hardware side, a number of application-specific integrated circuits (ASIC) solutions in CMOS[11–14] have been presented to help bring deep learning algorithms to a low-power processor. However, limitations still exist on memory footprint, static power consumption and in-memory computing. In particular, CMOS ASIC designs show that memory is the biggest bottleneck for energy-efficient real-time computing[11], in terms of storing millions of parameters and communicating them to the place where computing actually occurs. Although SRAM technology has been following the CMOS scaling trend well[15], the SRAM density (~150 $F^2$ per bitcell, F is the feature size of a technology node) and on-chip SRAM capacity (a few MB) are insufficient to hold the extremely large number of parameters in DNNs (even with binary precision), leakage current is undesirable, and parallelism is limited due to row-by-row operation[16]. To address the limitations of row-by-row operation, SRAM-based in-memory computing designs have been recently proposed, where multiple or all rows are turned on simultaneously and MAC computations of DNNs are performed along the bitlines with analog current or voltage[17–20]. This enhances parallelism for neural computing, but several additional transistors are typically added per bitcell in these designs, which aggravates the density issue as well as the leakage problem.

As an alternative hardware platform, emerging resistive devices with high on/off ratio have been proposed for weight storage and fast parallel neural computing with low power consumption[21,22]. The parallelism property of the resistive crossbar arrays for matrix-vector multiplication enables significant acceleration of core neural computations[23–28]. However, severe limitations still exist for applying resistive devices or resistive random access memories (RRAMs) for practical large-scale neural computing due to (1) device-level non-idealities, e.g., non-linearity in programming weights, variability, selector requirement, and endurance, (2) inefficiency in representing/multiplying negative weights and neurons, and (3) monolithic integration of RRAM arrays and CMOS peripheral read/write circuits. Due to these limitations, the RRAM based neural network hardware in the literature have mostly implemented simpler fully-connected multi-layer perceptrons (MLPs)[24,26], with very limited implementation of mainstream convolutional neural networks (CNNs) or recurrent neural networks (RNNs). Performing convolution operations was demonstrated using RRAM crossbar[25], but the application was limited to image processing/compression tasks, instead of deep neural networks. In addition, most RRAM hardware have been demonstrated without proper peripheral circuitries monolithically integrated in the same technology[24], and more often than not, peripheral circuits can dominate the chip area[29]. An RRAM macro with multi-level sense amplifiers in 55nm CMOS logic process was recently reported[30], but a relatively low accuracy of 81.83% accuracy for CIFAR-10 dataset was achieved with binary/ternary precision, and only 9 WLs are asserted simultaneously in the 256×512 sub-array, which limits further parallelism.

In this work, we address such limitations in RRAM based neural computing. We adhere to binary RRAM devices (low/high resistance states with high on/off ratio) and one-transistor-one-resistor (1T1R) structure for robustness against noise/variability and ease for integration. Using binary RRAM devices, we present new RRAM bitcell and array designs that can efficiently map XNOR functionality with binarized (+1, -1) weights/neurons and are suitable for in-memory computing that implements binarized DNNs. Using Winbond's commercial RRAM technology[31], we monolithically integrated a 128×64 RRAM array with CMOS peripheral circuits including row and column decoders/drivers and flash ADCs, which collectively can be a core component for large-scale RRAM-based in-memory computing. Based on measurement results from the prototype chips, we demonstrate deep CNNs for CIFAR-10[32] dataset as well as MLPs for MNIST dataset[33] with high classification accuracy and energy-efficiency.

## Results

**XNOR-RRAM prototype chip design.** We designed the custom RRAM array with robust in-memory computing operations using voltage-mode sense amplifier based flash ADC, and fabricated a prototype chip with Winbond's embedded RRAM technology[31], which monolithically integrates 90nm CMOS and RRAM between M1 and M2 (Fig. 1a). This is a substantial expansion beyond our preliminary simulation study[28], where binarized neural networks are mapped onto RRAM arrays with ideal RRAM device models with variability-prone current-mode sense amplifiers[34]. The pad-limited prototype chip micrograph is shown in Fig. 1b, and Fig. 1c shows the core area of the XNOR-RRAM chip. As shown in the top-level block diagram in Fig. 1d, the testchip design includes a 128×64 1T1R array, row decoder, level shifter, eight 8-to-1 column multiplexers, eight 3-bit flash ADCs, and two 64-to-1 column decoders for RRAM cell-level programming. The row decoder has two modes of operation: (1) it asserts all differential wordline (WL) signals simultaneously for XNOR-and-accumulate (XAC) operation, or (2) generates one-hot WL signals for cell-level programming. Eight ADCs (shared among 64 columns) and eight column multiplexers occupy 20% and 12% area of the XNOR-RRAM core, respectively (Fig. 1e).

**XNOR-RRAM bitcell design for BNNs.** Conventional binary RRAMs cannot effectively represent the positive and negative weight values (+1 and -1) in recent BNNs[9,10], because the high resistance state (HRS) and low resistance state (LRS) values of binary RRAM devices are both positive. In addition, as shown in Fig. 2c, the activation/weight value combinations of +1/+1 and -1/-1 should result in the same effective resistance. To that end, we proposed to use a "XNOR-RRAM" bitcell design[28] for XNOR-Net[9] type of BNNs. As shown in Fig. 2a, the XNOR-RRAM bitcell involves differential RRAM cells and differential wordlines. The binary activations are mapped onto the differential wordlines, and the binary weights are mapped onto the HRS/LRS values of XNOR-RRAM bitcells. By asserting all differential WLs of the RRAM array simultaneously, all cells in the same column are computed in parallel, which implements the XAC computations. The 128×64 1T1R array effectively represents 64×64 XNOR-RRAM bitcells, since one XNOR-RRAM bitcell consists of two 1T1R baseline bitcells to represent positive/negative weights and to perform embedded XNOR computation inside the XNOR-RRAM bitcell. The area of the 1T1R bitcell that we used is ~0.5μm*0.5μm (~31 $F^2$, where F is feature size), and thus one XNOR-RRAM bitcell occupies ~62 $F^2$ area.

While other recent embedded RRAM works also similarly employed separate RRAM cells for positive and negative weights[30,35,36], it is possible to use single cell for each BNN weight. Sequential row-by-row RRAM designs can use a single cell for +1/-1 weights and the binary MAC operation can be computed with dedicated digital logic on the periphery of the RRAM array. In our preliminary study[37], it has been shown that the sequential design has ~16% larger area due to such additional digital logic, compared to the in-memory computing based parallel RRAM design which embeds the binary MAC computation within the RRAM array. On the other hand, if the DNN activations do not have negative values (e.g. using ReLU activation), in-memory computing based RRAM designs could employ a single cell for each weight, while using a dummy column with the averaged LRS and HRS resistance (i.e., (LRS+HRS)/2), and subtracting the compute column's current from the dummy column's current. However, additional peripheral circuitry is needed for analog current subtraction (before ADC) or digital subtractor (using ADC output values). Overall, using two 1T1R cells enables true implementation of binarized DNNs with embedded XNOR operations, and also eliminates peripheral circuits that were necessary in prior works such as digital MAC or analog/digital subtraction.

On the other hand, while 6T SRAMs use aggressive push rules, in-memory computing SRAMs[17–20] require additional transistors beyond 6T. Until custom SRAM bitcells with additional transistors are engineered with push rules by foundry and semiconductor companies (Intel, TSMC, etc.), logic rules will need to be used for in-memory computing SRAM bitcells, which make the bitcell size even larger (~426 $F^2$ for 8T1C bitcell[20], ~927 $F^2$ for 12T bitcell[19]) Therefore, the density benefit of XNOR-RRAM using foundry RRAM can be still maintained, if we compare SRAM and RRAM both for in-memory computing.

**In-memory computing operation.** A static PMOS header, the strength of which is configurable, pulls up the RBL voltage. The RRAM cells in the same column pull down the RBL voltage in parallel. Depending on how many cells with high WL voltage are in LRS or HRS, a static resistive divider is formed between the PMOS head and the pull down path based on the parallel RRAM cells. As more RRAM cells are in LRS (higher bitcount value from the algorithm), BL voltage will be lower. The prototype chip measurement results with different PMOS header strengths are shown in Fig. 2e, where it can be seen that a stronger PMOS header increases the RBL voltage for the same bitcount value. We found that best accuracy is achieved with transfer curves that have the steepest slope around bitcount value of 0, since the flash ADC reference voltages could be separated further compared to the cases where the transfer curves have more gradual slope. Therefore, we chose to use the medium-strength PMOS configuration of 4 or 5 (in Fig. 2e) for our DNN workloads. For PMOS strength of 4 that we used for a prototype chip, the reference voltage values of the 7 VSAs for the 8 different ADCs are reported in Supplementary Fig. 2b.

Although high BL voltage (up to ~1V in Fig. 2e) can cause read disturb issues in RRAM cells[38], there are two things that largely prevent read disturb to occur in our XNOR-RRAM design. First, Fig. 2e shows that relatively high RBL voltage of >0.6V only occurs for small bitcount values that are lower than -32. On the other hand, the bitcount value distribution in Supplementary Fig. 1 shows that there is only <0.046% data in this range. Second, we experimentally observed that RRAM cells whose HRS resistance is larger than 1MΩ are stable, and are not susceptible to read disturb issues even with high BL voltages of >0.6V. On the other hand, we did observe that the outlier HRS cells with <1MΩ resistance can experience read disturb with high BL voltages for read operation. However, it can be found from Fig. 3a that only <1% of the programmed HRS cells exhibit less than 1MΩ resistance. Considering these two reasons and data distribution values above, the probability that read disturb will occur becomes extremely low (e.g. < 0.00046*0.01) in our XNOR-RRAM array read operation. Each 3-bit flash ADC consists of seven voltage-mode sense amplifiers (VSAs), whose outputs generate seven thermometer-coded bits that represent eight levels.

**ADC design and optimization.** Each VSA compares the read bitline (RBL) voltage of the selected column with a reference voltage. Seven reference voltages of an ADC are calibrated for the eight columns that the ADC is connected to. With the monotonic relationship between RBL voltage and bitcount values (Fig. 1e), the seven corresponding bitcount values we choose to serve as the reference points are: -13, -9, -5, -1, 3, 7, 11, out of the possible bitcount range between -64 and +64. These reference bitcount values are chosen in a confined range between -13 and 11, because the bitcount data distributions from DNN workloads are found to be highly centered around 0 (Supplementary Fig. 1). To verify the benefits of using the confined range for quantization, we performed software simulation with ideal quantization (no ADC offset, etc.) by using linear quantization for the full range of bitcount values from -64 to +64. For 3-bit, 4-bit, and 5-bit ADCs with "full-range" linear quantization, we obtained CIFAR-10 accuracies of 45.56%, 85.84%, and 88.59%. For 3-bit ADC with the proposed "confined-range" linear quantization, we achieved CIFAR-10 accuracy of 86.70% in software simulation with ideal quantization. It can be seen that 3-bit ADC with confined-range quantization achieves better accuracy than even 4-bit ADC with full-range quantization.

On the other hand, compared to nonlinear quantization schemes proposed in prior works[19,28], the proposed confined/linear quantization scheme simplifies the ensuing accumulation of ADC outputs (partial sums), and also increases the smallest reference voltage difference for adjacent sense amplifiers in the flash ADC. An automatic algorithm is employed to determine the reference voltages of the flash ADCs. For example, for a reference voltage corresponding to bitcount value of -13, we randomly generate 1,000 input vectors such that the resulted bitcount values are either -12 or -14. If the reference voltage is perfectly set for bitcount value of -13 in an ideal XNOR-RRAM array design, for all the input vectors that resulted in bitcount value of -12, the comparator output Q should be "0"; and for all input vectors that resulted in bitcount value of -14, Q should be "1". However, due to circuit non-ideal factors such as RRAM resistance variation, comparator noise, etc., misclassification could occur for both cases. The proposed algorithm aims to find the optimal set of comparator reference voltages that minimizes the misclassification (Supplementary Fig. 2a). The reference voltage is initially set at 0.6 V (half of VDD). After each of the 1,000 input vectors, the reference voltage is increased (or decreased if the correction amount is negative) by $\alpha\beta^n\times(Q_i - Q_a)$, where $Q_a$ is actual ADC output and $Q_i$ is ideal ADC output, $\alpha$ is initial correction step size (e.g., 5 mV), $\beta$ is a scaling factor (e.g., 0.995) which should be less than 1, and $n$ is the iteration index. If $Q_i = Q_a$ for a given input vector, no correction in reference voltage will be made in that iteration. The amount of correction decays exponentially such that the reference voltage will finally converge to a proper value that can discern the two adjacent bitcount values well.

To make a balance between area and throughput, we share one flash ADC by eight columns. In the functionality test mode, a 64-bit input vector is fed through a scan chain and the ADC outputs can be read out through the scan chain. In power measurement mode, random 64-bit input vectors are generated by linear-feedback shift register (LFSR) every eight cycles. The row decoder converts the 64-bit input vector to a 128-bit vector input to the XNOR-RRAM array by adding 64 complementary bits. Since the wordline voltages and the analog multiplexer gate voltage can be as high as 2-5V, which is higher than the CMOS standard-cell based logic level (1.2V), thick-oxide transistor (IO device) is used for the 1T1R bitcell and a level shifter is included as the driver for the row decoder and column decoder.

**LRS and HRS programming.** Fig. 3 shows the testchip measurement results of LRS and HRS distribution for the 128×64 array. Tightening LRS distribution is more important for our application of mapping BNNs onto the XNOR-RRAM array, because the column current will be dominated by current through LRS cells. We set the target of the LRS resistance value range to 5.9-6.1 kΩ. To achieve this, we apply an aggressive write verify scheme: First, we set the initial gate voltage to 2.3 V and apply a 100-ns SET pulse with amplitude of 2.1 V. If the resistance after SET is lower than the lower bound, i.e., 5.9 kΩ, a 200-ns RESET pulse with amplitude of 3.8 V and gate voltage of 4 V is applied to the RRAM cell followed with a SET pulse with a 0.05 V lower gate voltage; if the resistance after SET is higher than the upper bound, i.e., 6.1 kΩ, a RESET pulse is applied to the RRAM cell followed with a SET pulse with a 0.05 V higher gate voltage. We repeated the previous steps for up to 10 times until the LRS resistance falls in the target range. As for HRS resistance values, we set the target HRS resistance value to be above 1 MΩ. To achieve this range, we apply a 200-ns RESET pulse with amplitude of 3.8 V and gate voltage of 4 V to the RRAM cell and repeat applying the same RESET pulse up to 10 times until the resistance value is greater than 1 MΩ. In Supplementary Fig. 3, we show how the LRS and HRS distributions changed after iterative write-verify-read operations. Less than 1% HRS resistance values are lower than 1 MΩ; less than 10% LRS resistance values are lower than 5.9 kΩ; less than 25% LRS resistance values are greater than 6.1 kΩ; More than 99% LRS resistance values are in the range of 5.7-6.3 kΩ. The resistance values are read at 0.2 V by a source measurement unit (SMU). Although we go through up to ten times of SET/RESET operations for the initial programming, since we will not re-program the weights often for DNN inference applications, the endurance of >$10^5$ cycles reported by Winbond[31] is sufficient.

**Chip measurement results on in-memory computing.** In Fig. 2e, the measurement results of a single column is shown for RBL voltage against ideal bitcount values. The RBL voltage needs to go through the ADC. We explored three different reference voltage schemes for the flash ADC: (1) one set of unified reference voltages for the entire 8 ADCs of the testchip, (2) 8 sets of reference voltages for 8 ADCs (one set per ADC), and (3) 64 sets of reference voltages for 64 columns (one set per column). For these three schemes, in Fig. 4, we show the comparison of the bitcount values from the binarized DNN algorithm (ideal partial sum value) and the measured ADC output values. We also show the comparison between the ideal ADC output and measured ADC output. It can be seen that the bitcount value and the ADC output show an expected linear relationship (highlighted in bright color). We first programmed the XNOR-RRAM with a 64×64 weight submatrix from the trained binary neural network for MNIST using aforementioned write-verify scheme; 2,000 64-bit binary test vectors were then presented to XNOR-RRAM, to perform XNOR-and-accumulate computations and obtain the 2,000×64 ADC outputs; In total, 128,000 pairs of measured ADC outputs and target XAC bitcount values are used to estimate the joint distribution of these two. The 2-D histograms in Fig. 4 shows how accurately the XNOR-RRAM array computes and quantizes the XAC values. Fig. 4a and 4b shows that using only one set of unified reference voltages without offset calibration can result in large variations in the ADC output. However, if each ADC has its own reference voltages (Fig. 4c and 4d), or exhibits offset cancellation capability, then the ADC output resides in a tight range for each bitcount value. If each column has its own reference voltages (Fig. 4e and 4f), it can been seen that there is not only minor difference compared the results in Fig. 4c and 4d. As an initial prototype chip design, please note that our VSA and ADC design did not include offset compensation circuits or techniques. If our VSA/ADC had employed offset cancellation circuits typically accompanied in sense amplifier designs[39], then all ADCs in our XNOR-RRAM macro would be able to use the same reference voltages, enabling the in-memory computing design with higher practicality.

**DNN accuracy characterization.** With these three schemes, we benchmarked the accuracy of the proposed XNOR-RRAM array for large deep neural networks for MNIST and CIFAR-10 datasets (Fig. 5a). For MNIST, we used a multilayer perceptron (MLP) with a structure of 784-512-512-512-10. For CIFAR-10, we employed a convolutional neural network (CNN) with 6 convolution layers and 3 fully-connected layers[10]. Supplementary Fig. 4 illustrates how fully-connected and convolution layers of MLP and CNN are mapped onto multiple XNOR-RRAM instances, where weights for different input channels are stored on different rows, weights for different output channels are stored on different columns, and weights within each convolution kernel (e.g. 9=3x3) are stored in different XNOR-RRAM macros. Subsequently, the partial MAC results from different XNOR-RRAM macros are accumulated via digital simulation. Fig. 5b and Fig. 5c shows the measurement results with the three different reference voltage schemes for MNIST MLP and CIFAR-10 CNN, respectively. Compared to the scheme with a single set of reference voltages for the ADC (without offset calibration), using 8 sets of reference voltages for 8 ADCs show noticeable improvement in both MNIST and CIFAR-10 accuracy values. This would be largely due to the local mismatch of the transistors in the ADC, which can be compensated by offset cancellation schemes typically employed in ADC designs[39]. On the other hand, the accuracy values for the scheme using 64 sets of reference voltages is hardly different to those using 8 sets of reference voltages. This means that, while offset cancellation for ADCs is important, column-by-column variation is not large and does not affect the accuracy noticeably. Using 8 sets of reference voltages, XNOR-RRAM achieves 98.5% classification accuracy for MNIST dataset (software binary MLP baseline: 98.7%), and achieves 83.5% classification accuracy for CIFAR-10 dataset (software binary CNN baseline: 88.6%). The accuracy degradation for CIFAR-10 dataset occurs due to limited ADC precision and small separation in adjacent reference voltages of ADC (caused by the gradual slope of bitline transfer curve), which could be improved by employing an ADC with higher precision[40] (trading off ADC area and power) or asserting less number of rows[30] in parallel to reduce the dynamic range (trading off latency or energy-efficiency).

**Power, energy and throughput results.** We measured the power of the prototype chip under different power supply voltages (1.2 V down to 0.9 V) for the PMOS pull-up and ADC. As we lower the PMOS pull-up power supply voltage, the current of the voltage dividers decreases, reducing the total power and improving the energy efficiency as shown in Supplementary Fig. 5. However, as we reduce the power supply voltage, the ADC sensing margin greatly reduces, degrading the accuracy on BNN benchmarks. For example, for MNIST MLP, the accuracy degrades to 97.28% when power supply voltage is 0.9 V; and for CIFAR-10 CNN, the accuracy degrades to 80.65% when power supply voltage is 1.1 V. On the other hand, as we decrease the operation frequency, the throughput decreases accordingly, however, the current of voltage dividers does not change much, which worsens the energy efficiency at low operation frequency (Supplementary Fig. 5). Therefore, to fully take advantage of the XNOR-RRAM array, we should run prototype chip at its highest possible operation frequency, which is governed by the RBL settling time. Due to measurement issues with on-chip generated clocks, the clock frequency for chip measurement was limited by the slow IO pads at ~20MHz. To that end, we report post-layout extracted simulation results for fastest operating frequency for the XNOR-RRAM chip. As shown in Supplementary Fig. 6, the critical path from clock rising edge to the RBL settling time is 6.5 ns, which means that the XNOR-RRAM prototype chip could operate at 154 MHz, achieving energy efficiency of 24.1 TOPS/W (Tera operations per second per Watt) and 29.2 TOPS/W at 1.2 V and 1.1 V power supply, respectively.

With all 128 rows and 8 columns (8 ADCs shared among 64 columns) asserted and computed simultaneously in each cycle, the 128×64 XNOR-RRAM array achieves a high throughput of 157.7 GOPS. Supplementary Fig. 7 shows the comparison with the closest prior work implemented in 55nm CMOS with embedded RRAM[30]. For the same binary precision, our work achieves higher CIFAR-10 accuracy. Xue et al.[30] only turns only 9 rows simultaneously for in-memory computing, while our work turns on all 128 rows in the sub-array, leading to 5.6X higher throughput per ADC operation. For in-memory computing, turning on more rows typically require ADCs with higher precision due to higher dynamic range of MAC results. However, we turn on all 128 rows and achieve better binary-DNN accuracy than Xue et al.[30] even with lower-precision ADC (3-bit ADC in our work vs. 4-bit ADC in Xue et al.[30]), aided by both the confined-range linear quantization (Supplementary Fig. 1) and the ADC reference voltage optimization (Supplementary Fig. 2a).

We considered two figure-of-merits (FoMs), FoM1 and FoM2 that represent two well-known energy/performance metrics for computer systems, energy-delay product (EDP) and energy-delay-squared product ($ED^2P$), respectively. FoM1 is the product of energy-efficiency (TOPS/W) and throughput per ADC (GOPS), effectively representing the inverse of energy-delay-product (EDP), which is a well-known metric that reflects a balance between energy and performance. FoM2 is the product of energy-efficiency (TOPS/W) and the square of throughput per ADC ($GOPS^2$), effectively representing the inverse of energy-delay-squared product ($ED^2P$), which is known as a more appropriate metric that presents energy and performance trade-offs with voltage scaling[41]. Our work achieves 3.2X higher FoM1 and 14.1X higher FoM2, compared to those of the state-of-the-art literature[30]. These improvements will be even higher if we normalize the CMOS technology (55nm[30] versus 90nm for our work). Exhibiting high throughput and low EDP/$ED^2P$ is essential for performance-critical or real-time operating systems (e.g. autonomous driving, real-time machine translation), and our in-memory computing technique becomes very suitable for such latency-constrained artificial intelligence systems.

## Discussion

We demonstrated an energy-efficient in-memory computing XNOR-RRAM array, which turns on all differential wordlines simultaneously and performs analog MAC computation along the bitlines. By monolithically integrating flash ADCs and 90nm CMOS peripheral circuits with RRAM arrays, we demonstrate the scalability of XNOR-RRAM towards large-scale deep neural networks. XNOR-RRAM prototype chip measurements and extracted simulations demonstrate high energy-efficiency of 24 TOPS/W, high throughput of 158 GOPS, and high classification accuracy of 98.5% and 83.5% for MNIST and CIFAR-10 datasets, respectively. Our work achieves 3.2X improvement in EDP and 14.1X improvement in ED$^2$P, compared to those of the state-of-the-art literature.

ADCs generally incur a large overhead for in-memory computing especially with dense NVMs, as also reported by prior works[42,43]. Unlike SRAM, the RRAM column pitch is less, which makes the core even more dominated by the peripheral circuits. We used flash ADC where area will exponentially increase with bit precision, therefore, a possible trade-off is to use more compact successive-approximation-register (SAR) ADC[30,44], while allowing longer latency, in order to reduce the area overhead.

In our XNOR-RRAM design, further energy-efficiency improvement is largely governed by the LRS resistance of the RRAM technology. Higher energy-efficiency could be achieved by RRAM technologies with higher LRS resistance values, while this consequently will reduce the on/off ratio. Our current XNOR-RRAM only supports binarized DNNs (both activations and weights have +1 or -1 values), but multi-bit precision DNNs that lead to higher accuracy could be supported by bit-serial operation and additional digital peripheral circuits[45] and/or digital-to-analog (DAC) converters[40], while sacrificing energy-efficiency. However, the core in-memory computing technology with the proposed 2T2R bitcell design and peripheral ADC can be applied generally to any given RRAM technology and RRAM arrays.

## Methods

**Experimental design.** We designed and fabricated the XNOR-RRAM prototype chip using Winbond's 90nm CMOS technology. The 128×64 1T1R bitcell array was provided by Winbond. We designed the column multiplexer and flash ADC and laid out manually with transistors in 90nm CMOS technology. The wordline/column decoder, level shifter, and scan chain modules were designed by Verilog, synthesized using Synopsys Design Compiler tool, and automatically placed and routed using Cadence Innovus tool, following the typical digital integrated circuit design flow with 90nm CMOS standard cell libraries. The final layout was sent to Winbond and the prototype chip has been fabricated with Winbond's embedded NVM technology, which monolithically integrates RRAM and 90nm CMOS.

**Electrical characterization.** The prototype chips were assembled in QFN80 packages. The prototype chip sits in a QFN80 test socket on a custom testing PCB, which connects to several National Instruments testing modules including NI PXIe-5413 for programming pulse generation, NI PXIe-6555 for digital signal generation and acquisition, NI PXIe-4140 for RRAM cell resistance measurement, NI PXIe-6738 for reference voltage and gate voltage generation and NI TB-2630B for analog signal multiplexing (Supplementary Fig. 8).

**Prototype chip measurement.** We used on-chip decoder to select a single 1T1R cell such that we could do forming/SET/RESET/read operation for each cell independently. We first applied forming pulses (20 µs, 3.8 V) with gate voltage at 1.9 V to the RRAM array (128×64). Depending on the target weight sub-matrix (64×64), we programmed each cell of the array to either high resistance state (HRS) or low resistance state (LRS). To change a cell from HRS to LRS, we applied SET pulses sequentially. After each SET pulse, the cell resistance is read out and we stopped applying SET pulses if it reached the specified LRS target range (e.g., 5.9 kΩ to 6.1 kΩ); if the resistance is lower than the target range (e.g., less than 5.9 kΩ), we applied a RESET pulse followed by a SET pulse with the gate voltage decreased by 50 mV; if the resistance is greater than the target range (e.g., greater than 6.1 kΩ), we continued applying a SET pulse with the gate voltage increased by 50 mV. The SET pulses are of 100 ns pulse width and 2.1 V pulse amplitude. The initial gate voltage for SET operation is 2.3 V. To change a cell from LRS to HRS, we applied RESET pulses repeatedly (up to 10 pulses) until the resistance is greater than the specified HRS target resistance (1 MΩ). The RESET pulses are of 200 ns pulse width and 4 V (opposite polarity) pulse amplitude. The gate voltage is set at 3.8 V during RESET operation. After the weight sub-matrix was programmed to the RRAM array, we switched the chip into XNOR read mode. For functionality verification, we could scan in 64-bit input vectors through, and scan out the ADC outputs and compare them with ideal ADC outputs. We scanned in 2,000 test input vectors from MNIST BNN benchmark, and obtained a 2-D histogram of pairs of XNOR-accumulate (XAC) bitcount values and measured ADC outputs as shown in Fig. 4, from which we could get a conditional probability distribution of ADC outputs as a function of XAC bitcount values. For power measurement, we could generate random 64-bit input vectors internally on the chip through a linear-feedback shift register (LFSR). The vectors are updated every 8 cycles as the ADCs are shared by eight columns.

**Large DNN evaluation.** Due to the limitation of cell-by-cell programming on a single array using our test equipment, it would take months to complete all the operations for a deep binary neural network (BNN) such as a convolutional BNN for CIFAR-10 dataset. To speed up evaluation on such deep BNN, we ran behavioral software emulation based on the conditional probability distribution of ADC outputs on XAC bitcount values we measured from 2,000 random test vectors. In our software emulation, the deep convolutional kernels or weight matrices are divided in 64×64 weight sub-matrices. The partial sums from product between 64-bit inputs and 64×64 weight sub-matrices are first stochastically quantized to 3-bit according to the measured conditional probability distribution and then accumulated to produce the final products followed by element-wise operations such as batch normalization, max pooling, activation binarization, etc. We evaluated an MLP BNN for MNIST dataset and a convolutional BNN for CIFAR-10 dataset in this way, each with 20 runs with different random seeds. The accuracy numbers are summarized in box plots in Fig. 5b and Fig. 5c.

Programmable In-Memory Vector Acceleration. in *Proceedings of the IEEE International Solid-State Circuits Conference (ISSCC)* (2019). doi:10.1109/ISSCC.2019.8662419

**Acknowledgements**

We are grateful for chip fabrication support by Winbond Electronics. This work is partially supported by NSF-SRC-E2CDA under Contract No. 2018-NC-2762B, NSF grant 1652866, NSF grant 1715443, NSF grant 1740225, JUMP C-BRIC and JUMP ASCENT (SRC program sponsored by DARPA). We thank Prof. J. Kim for technical discussion. This work was performed at Arizona State University.

**Author Contributions**

J.S. and S.Yu conceived this work and directed the team. S.Yin designed the digital peripheries and executed top-level chip integration. X.S. designed the flash ADC and column multiplexers. S.Yin and X.Sun carried out prototype chip measurements and characterization. S.Yin performed the post-layout simulation work. S.Yin, S.Yu and J.S. prepared the manuscript. All authors discussed and contributed to the discussion and analysis of the results regarding the manuscript at all stages.

**Additional Information**

Correspondence and requests for materials should be addressed to J.S. (jaesun.seo@asu.edu) and S.Yu (shimeng.yu@ece.gatech.edu).

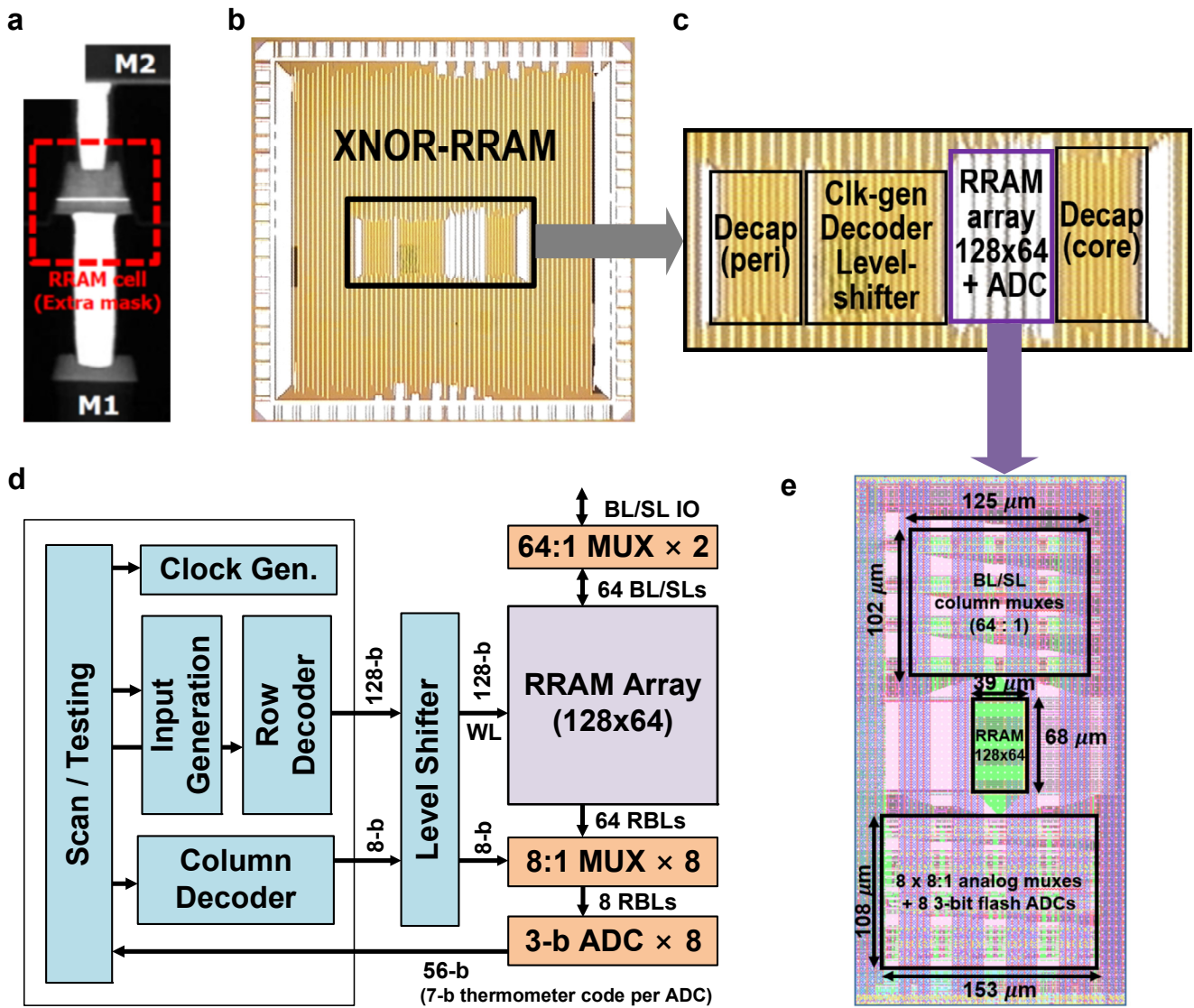

**Figure 1. XNOR-RRAM prototype chip design fabricated with monolithically integrated RRAM and 90nm CMOS technology. a.** Winbond's technology integrates RRAM between M1 and M2. **b.** The pad-limited prototype chip micrograph. **c.** The prototype chip's core area is shown, consisting of RRAM array and ADC, decoder and level-shifters, and decoupling capacitors. **d.** Our XNOR-RRAM chip design that integrates decoder periphery, 1T1R RRAM array, column multiplexers, and eight 8-level flash ADCs. **e.** Layout and dimensions of the RRAM array, column multiplexers and flash ADC are shown.

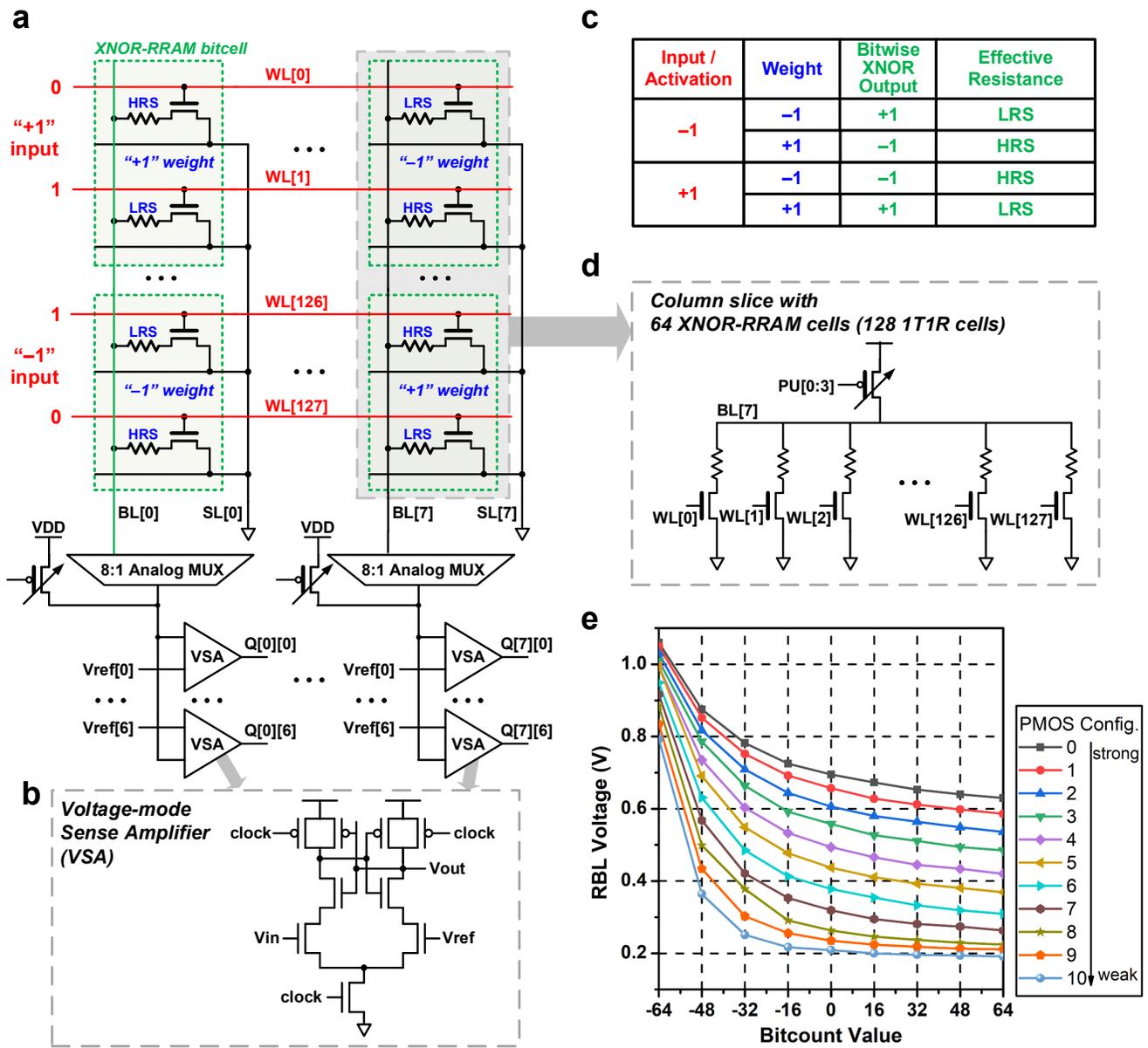

**Figure 2. In-memory computing operation of XNOR-RRAM a.** XNOR-RRAM bitcell consists of two 1T1R cells, where the differential programmed resistances represent DNN weights and the differential wordlines represent DNN inputs/activations. 3-bit flash ADC consists of seven voltage-mode sense amplifiers (VSAs). **b.** The schematic of the VSA is shown, which compares Vin and Vref at the rising edge of the clock. **c.** XNOR-RRAM bitcell embeds XNOR operation of inputs and weights, which governs the effective resistance. **d.** By turning on all 128 rows simultaneously, the column slice becomes a resistive divider between static PMOS header and the pull-down network with 64 XNOR-RRAM cells connected in parallel. **e.** The column measurement results with different PMOS header strengths show RBL voltage monotonically decreases with increasing bitcount values.

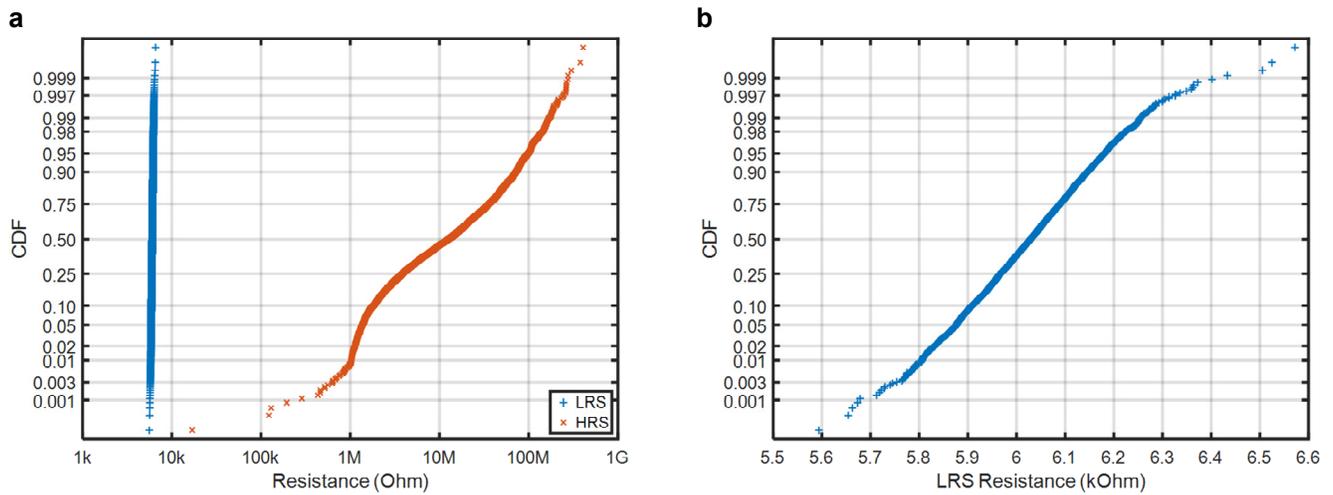

**Figure 3. The programming results and distributions of RRAM devices for XNOR-RRAM array. a.** Resistance distribution of 4,096 RRAM cells programmed in LRS and 4,096 RRAM cells programmed in HRS for a given 64×64 XNOR-RRAM array, where two complementary RRAM cells represent one binary weight. X-axis is in log scale and y-axis is in a fitted Gaussian cumulative distribution function (CDF) scale. **b.** LRS resistance distribution tightened to a confined range near 6 kΩ.

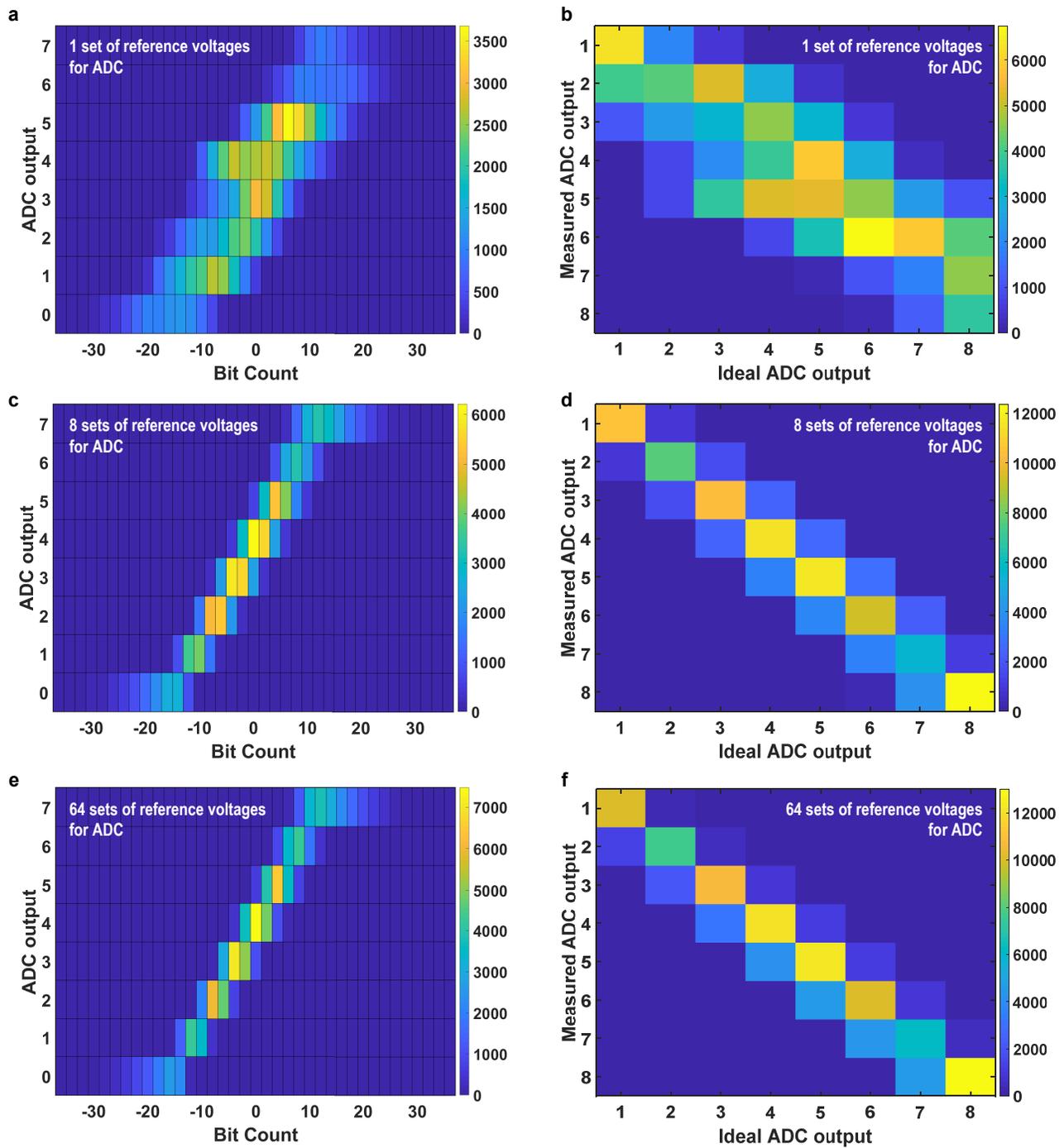

**Figure 4. Ideal partial sum results of binary MAC operations for DNNs are compared with measured ADC results.** From top to bottom: (a)(b) a single set of ADC reference voltages calibrated for all the 8 ADCs; (c)(d) Reference voltages of each ADC are calibrated separately; (e)(f) Reference voltages for each column are calibrated separately.

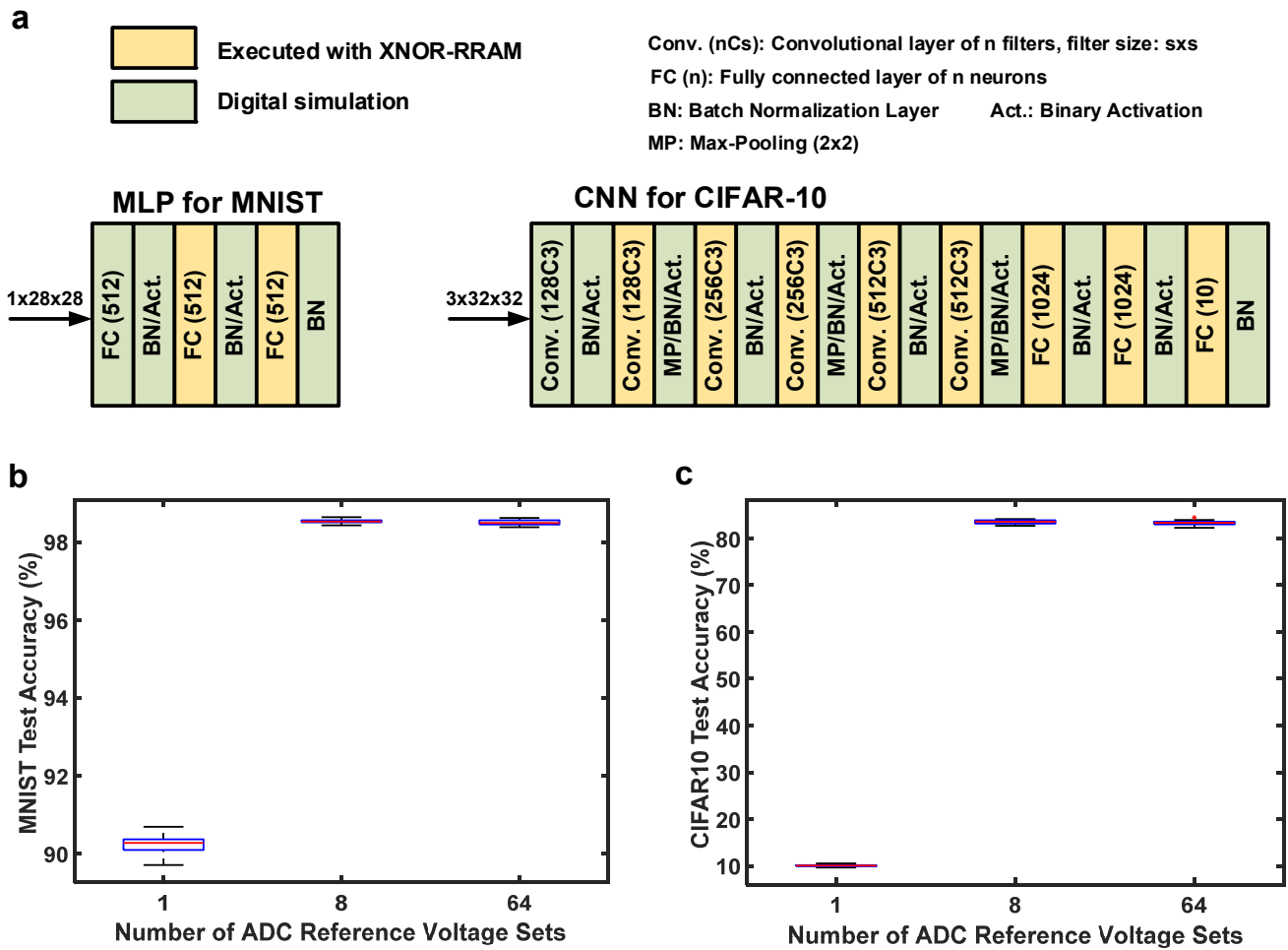

**Figure 5. Evaluation of DNNs for MNIST and CIFAR-10 datasets using XNOR-RRAM. a.** For MNIST, we used a multilayer perceptron (MLP) with a structure of 784-512-512-512-10. For CIFAR-10, we employed a convolutional neural network (CNN) with 6 convolution layers and 3 fully-connected layers. **b.** Accuracy numbers are obtained from 20 runs, where the partial sums in each run are stochastically quantized to 3-bit values, based on the probability distribution of the 2-D histogram in Fig. 4a, 4c, and 4e for 1, 8, and 64 sets of reference voltages for ADCs, respectively. The redline, box top edge, box bottom edge, top bar and bottom bar represents the mean, 75th percentile, 25th percentile, maximum and minimum of the 20 data points. The large accuracy increase from 1 set to 8 sets of reference voltages show that offset cancellation for the ADCs is important. The negligible accuracy difference between 8 sets and 64 sets of reference voltages represent that column-by-column variation due to LRS/HRS or interconnects do not affect accuracy noticeably.

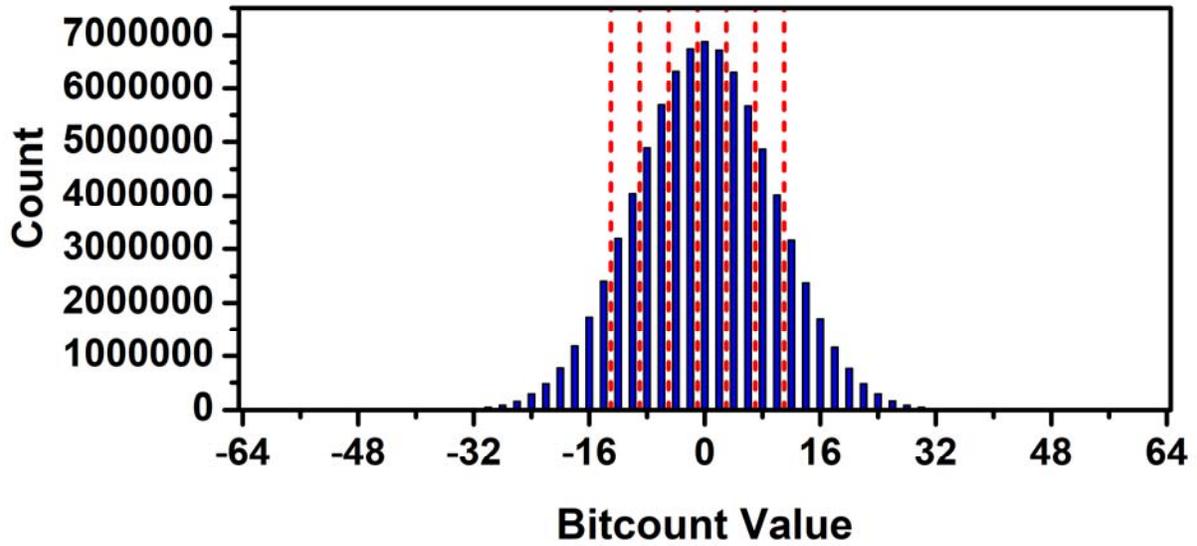

**Supplementary Figure 1. Bitcount value distribution from DNN workload.** By running the MLP for MNIST shown in Fig. 5, we characterized the distribution of ideal bitcount values that should be obtained from XNOR-RRAM arrays. Based on this distribution, we determined seven quantization edges at -13, -9, -5, -1, 3, 7, 11 (red dashed lines) for the 3-bit flash ADC. Exploiting the fact that there is scarce data near -64 and +64, we employ linear quantization within a confined range between -15 and 11.

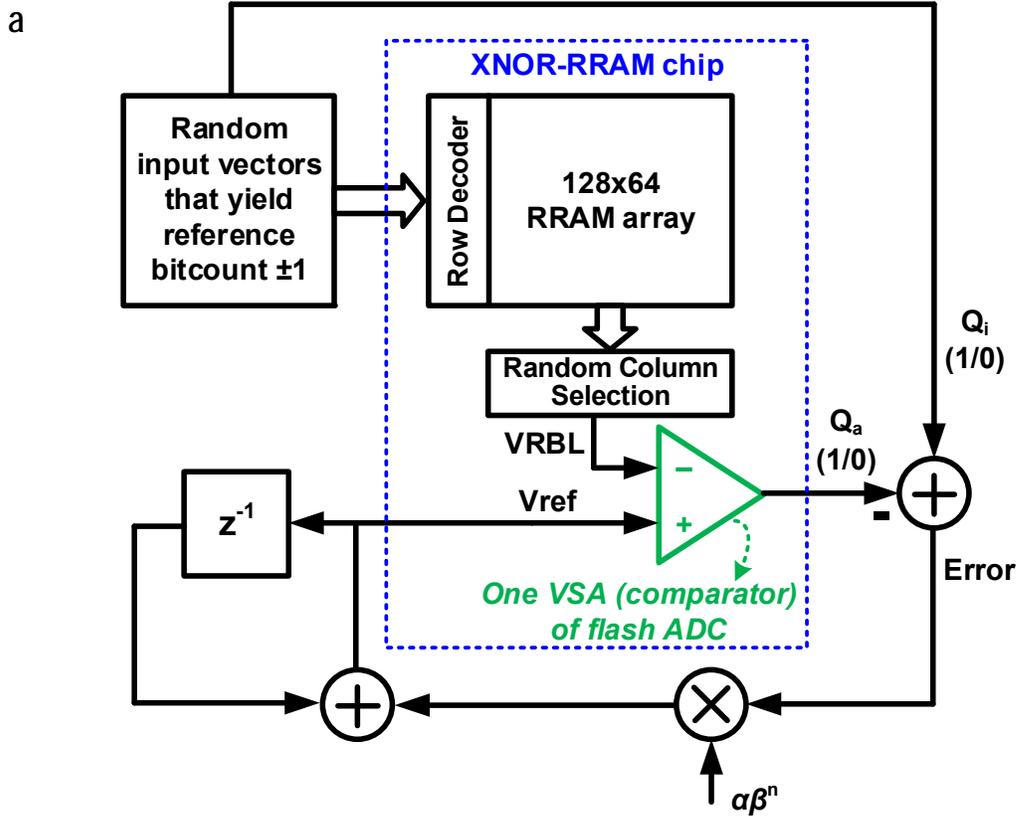

|        | VSA[0] | VSA[1] | VSA[2] | VSA[3] | VSA[4] | VSA[5] | VSA[6] |
|--------|--------|--------|--------|--------|--------|--------|--------|
| ADC[0] | 0.5335 | 0.5830 | 0.5553 | 0.5049 | 0.6121 | 0.5428 | 0.5559 |
| ADC[1] | 0.5566 | 0.5691 | 0.5822 | 0.6091 | 0.5366 | 0.5812 | 0.5326 |
| ADC[2] | 0.5309 | 0.5269 | 0.5722 | 0.5457 | 0.5249 | 0.5562 | 0.5611 |
| ADC[3] | 0.5744 | 0.5968 | 0.6254 | 0.5795 | 0.5599 | 0.5268 | 0.5359 |
| ADC[4] | 0.5557 | 0.5638 | 0.5473 | 0.5759 | 0.5391 | 0.5623 | 0.5687 |
| ADC[5] | 0.5669 | 0.6082 | 0.5946 | 0.5686 | 0.5747 | 0.5210 | 0.5511 |
| ADC[6] | 0.5415 | 0.5568 | 0.5941 | 0.5854 | 0.5795 | 0.5549 | 0.5770 |
| ADC[7] | 0.5701 | 0.5533 | 0.5570 | 0.5789 | 0.5771 | 0.5501 | 0.5901 |

**Supplementary Figure 2. ADC reference voltage calibration.**
**a.** For each comparator corresponding to a reference bit count, random input vectors that yield an output of reference bitcount ± 1 are fed to XNOR-RRAM chip, and the BL from a random column that is connected to the comparator is selected and BL voltage is compared with reference voltage. The reference voltage (Vref) is adjusted with exponentially decaying correction according to the comparison error. In particular, Vref is increased (or decreased if the correction amount is negative) by $\alpha\beta^n \times (Q_i - Q_a)$, where $Q_a$ is actual ADC output and $Q_i$ is ideal ADC output, $\alpha$ is initial correction step size (e.g., 5 mV), $\beta$ is a scaling factor (e.g., 0.995) which should be less than 1, and $n$ is the iteration index. If $Q_i = Q_a$ for a given input vector, no correction in reference voltage will be made in that iteration. Vref will finally converge to a proper value that can discern the two adjacent bitcount values well.
**b.** For PMOS strength of 4, the optimized Vref values of the 7 VSAs for the 8 different ADCs are listed for a protype chip that was calibrated. Note that Vref values are not monotonic from VSA[0] to VSA[6], because different VSAs exhibit different positive or negative offsets, and the Vref values have been calibrated against them.

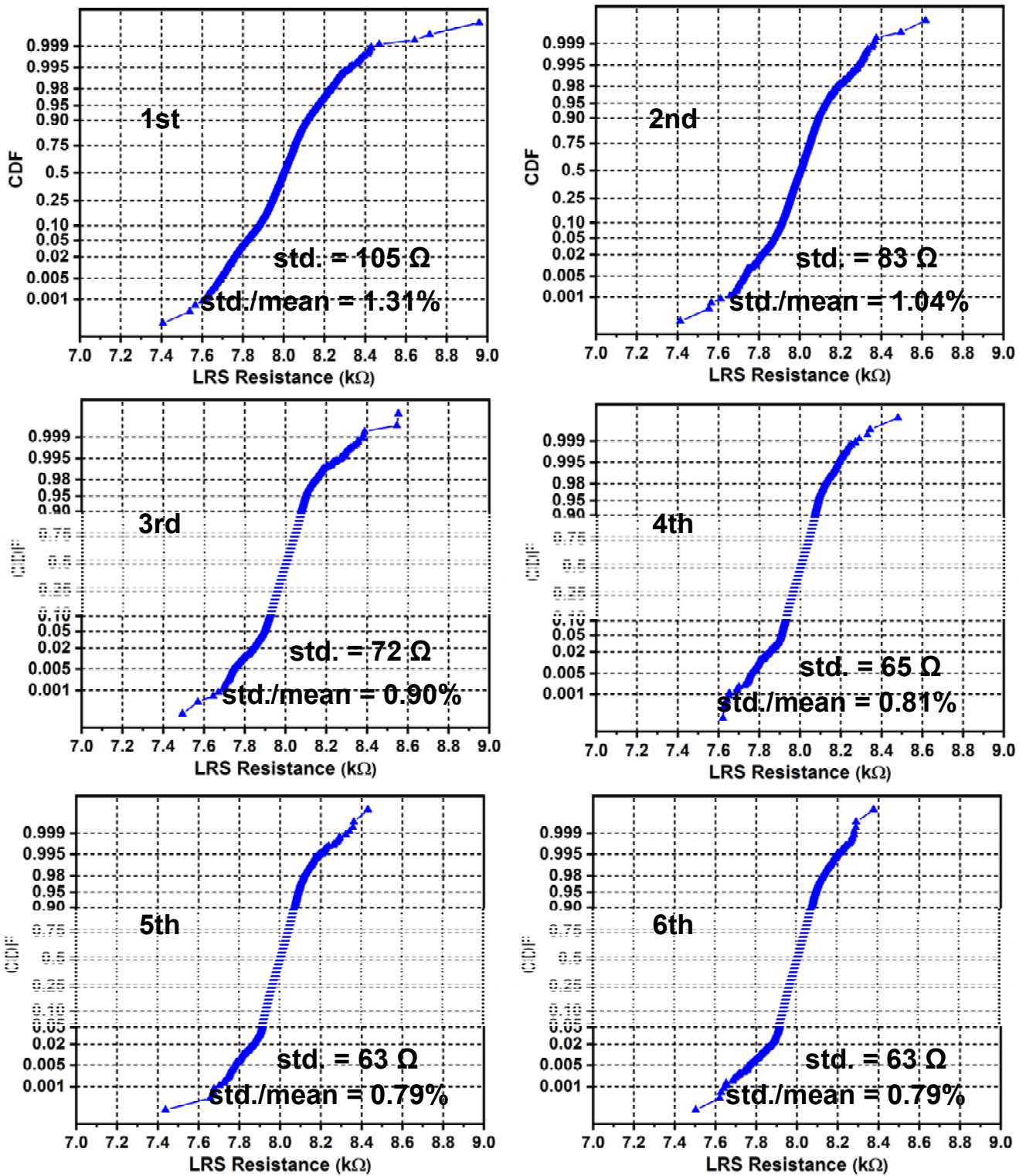

**Supplementary Figure 3. Programming iterations.** As we repeatedly program the RRAM array for the same weights, LRS resistance distribution becomes further tightened. From 1st to 6th programming iteration, the standard deviation (std.) of LRS resistance reduces from 105 Ω to 63 Ω, and the ratio of std. to mean reduces from 1.31% to 0.79%.

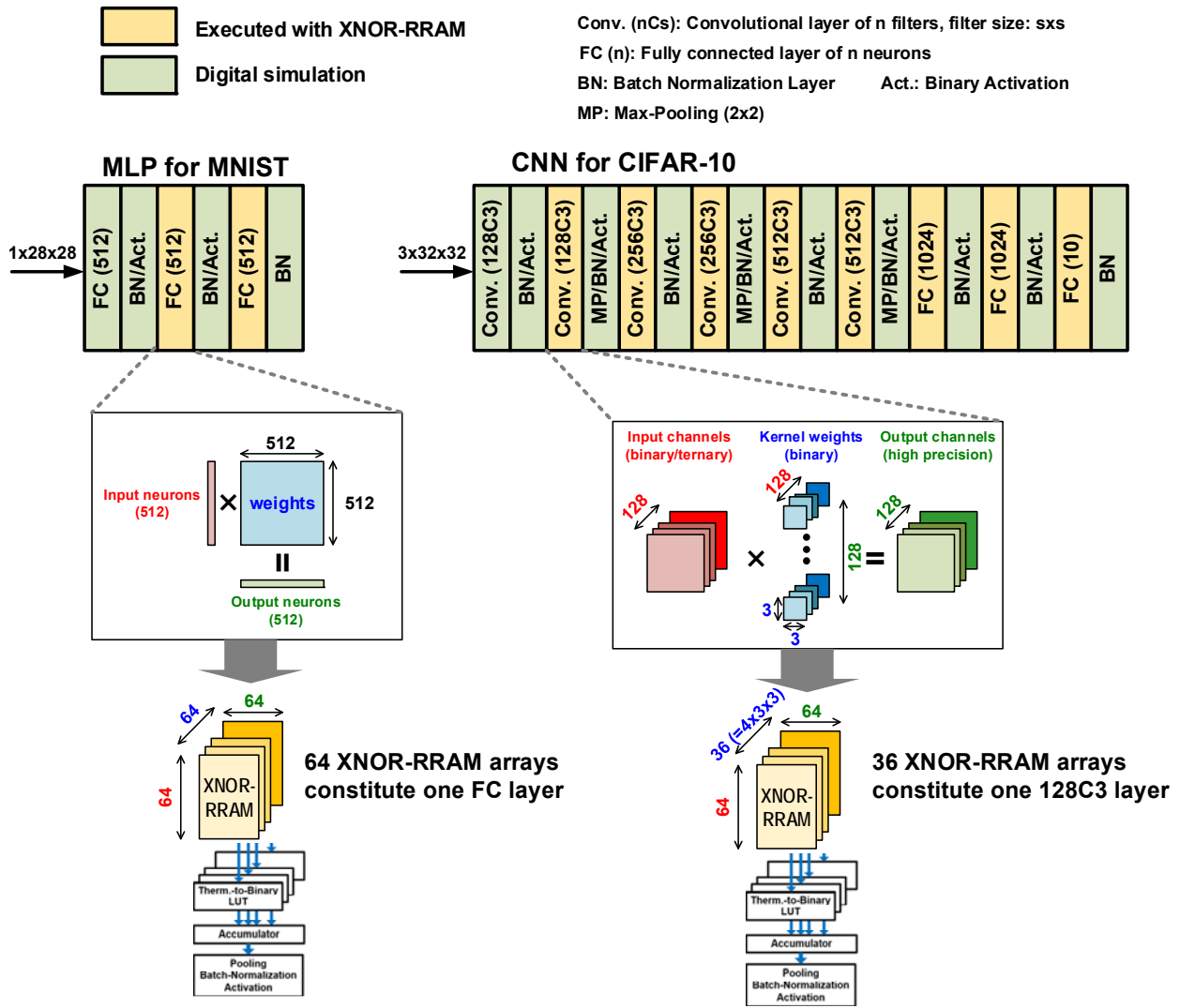

**Supplementary Figure 4. MLP and CNN evaluations.** An MLP BNN for MNIST and a convolutional BNN for CIFAR-10 are evaluated with XNOR-and-accumulate operations executed on XNOR-RRAM arrays and other element-wise operations executed on digital simulator with fixed-point precision. An FC layer of 512×512 weights is mapped to 64 XNOR-RRAM arrays. A 128C3 convolutional layer of 128×128×3×3 weights is mapped to 36 XNOR-RRAM arrays.

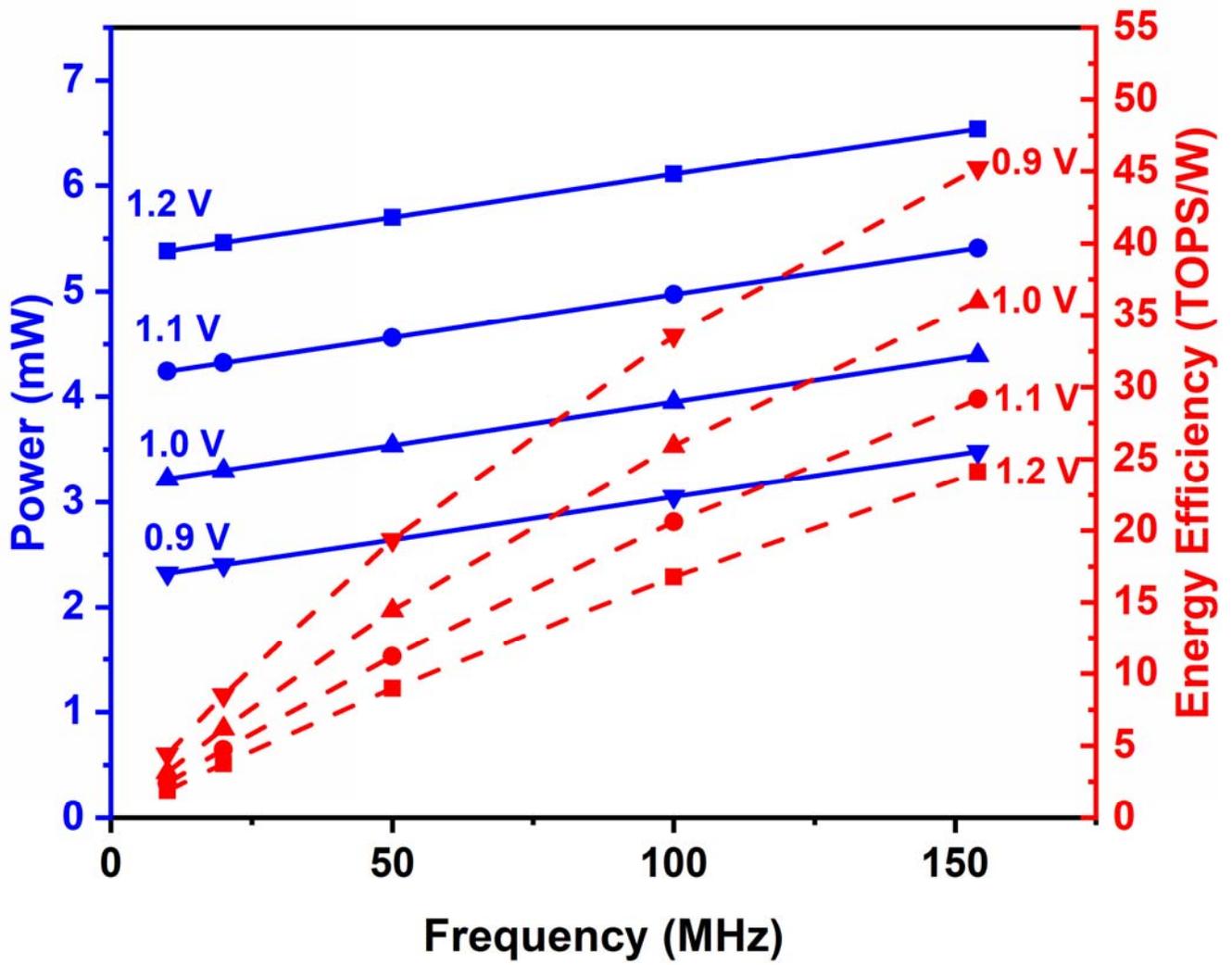

**Supplementary Figure 5. Power and energy measurements with voltage/frequency scaling.** For supply voltages from 1.2V down to 0.9V, we measured the power consumption of XNOR-RRAM prototype chip and characterized the energy efficiency (TOPS/W) at different frequency values.

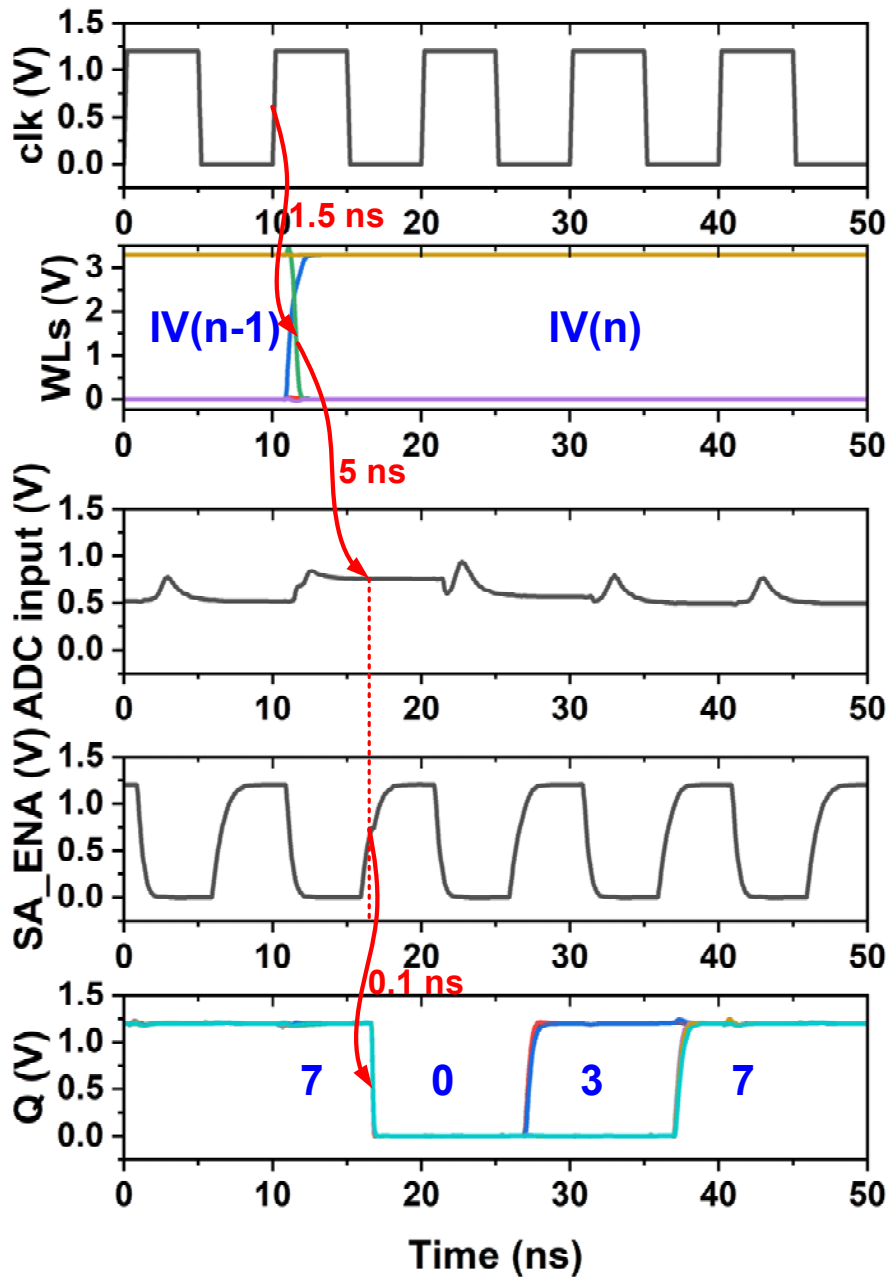

**Supplementary Figure 6. Extracted simulation results.** Each input vector is presented to XNOR-RRAM array for 8 cycles. Each ADC senses the RBL voltage of one of the 8 columns each cycle. From top to bottom: global clock signal, word lines driving the XNOR-RRAM array, RBL voltage as the ADC input, ADC clock signal (sense amplifier enable signal), flash ADC outputs. The clock to word line delay, RBL voltage settling delay, and flash ADC sensing delay are 1.5 ns, 5 ns and 0.1 ns, respectively.

|  | Xue et al.[30] | | This work |
| --- | --- | --- | --- |
| CMOS Technology | 55nm | | 90nm |
| Sub-array size | 256×512b | | 128×64b |
| Operating voltage | 1V (0.9-1.1V) | | 1.2V |
| # of rows turned on simultaneously | 9 | | 128 |
| # of operations per ADC operation | 36 | | 128 |
| Precision (bits) activation / weight / output | A:1 / W:ternary / O:4 | A:2 / W:3 / O:4 | A:1 / W:1 / O:3 |
| Energy-Efficiency (TOPS/W) | 53.17 | 21.9 | 24.1 |
| Read IMC delay (ns)[a] | 10.2 | 14.6 | 6.5 |
| Throughput (GOPS) per ADC operation[b] | 3.53 | 2.47 | 19.7 |
| FoM1 (Energy-Efficiency × Throughput per ADC)[c] | 149.1 | 72.8 | **475.3 (3.2X higher)** |
| FoM2 (Energy-Efficiency × Throughput$^2$)[d] | 662.5 | 133.6 | **9353.0 (14.1X higher)** |
| CIFAR-10 accuracy | 81.83% | 88.52% | 83.5% |

[a] Read IMC delay represents the delay to perform one in-memory computing (along the column) including the ADC operation.
[b] (# of operations) / (read IMC delay)
[c] This figure-of-merit (FoM1) effectively represents the inverse of energy-delay product (EDP), a well-known metric that balances energy and performance requirements.
[d] This figure-of-merit (FoM2) effectively represents the inverse of energy-delay$^2$ product (ED$^2$P), which is known as a more appropriate metric that presents energy and performance trade-offs with voltage scaling[35].

**Supplementary Figure 7. Comparison with prior work.** The performance of our XNOR-RRAM chip and the 55nm CMOS chip with embedded RRAM[30], is compared. For the same binary precision, our work achieves higher accuracy for CIFAR-10 dataset. Xue et al.[30] only turns only 9 rows simultaneously for in-memory computing, which adversely affects the throughput, and hence our work achieves 5.6X higher throughput per ADC operation. FoM1 and FoM2 effectively represent the inverse of EDP and the inverse of ED$^2$P, respectively. Our work achieves 3.2X improvement in FoM1 (EDP) and 14.1X improvement in FoM2 (ED$^2$P). These improvements will be even higher if we normalize the CMOS technology (55nm versus 90nm).

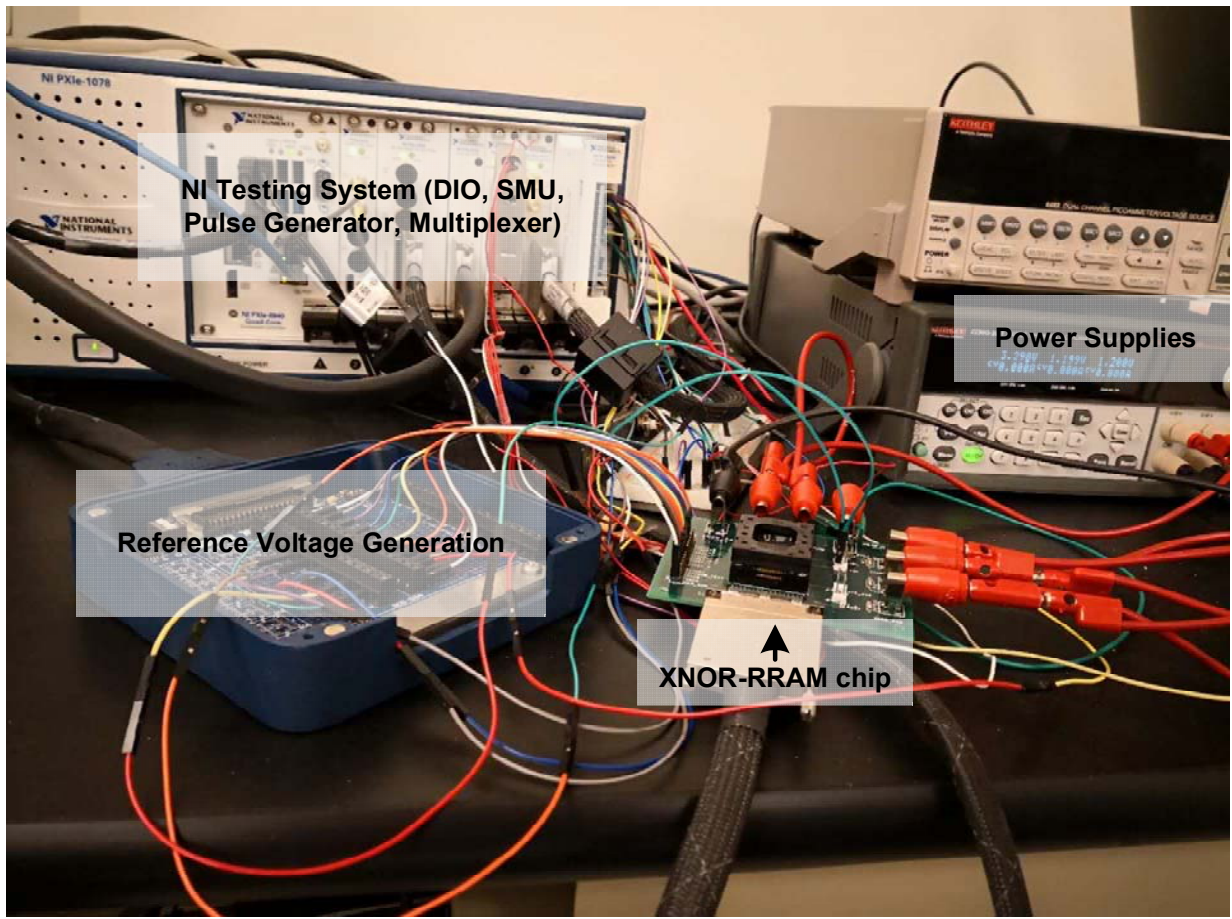

**Supplementary Figure 8. Testing setup photo.** The prototype XNOR-RRAM chip is connected to a print circuit board (PCB) via a testing socket. National Instruments (NI) testing system is employed to generate/acquire digital I/O signals, measure RRAM resistance through source measurement unit (SMU), generate SET/RESET pulses, multiplex BL/SL signals, and generate ADC reference voltages.